\documentclass[aip, amsmath,amssymb,reprint]{revtex4-1}
\usepackage{graphicx}
\usepackage{dcolumn}
\usepackage{bm}
\usepackage{graphicx}
\usepackage{xcolor}
\graphicspath{{./figures/}}
\usepackage{hyperref}
\usepackage{amsfonts}
\usepackage{amssymb}
\usepackage{amsmath}
\usepackage{notoccite}
\usepackage{lineno}
\usepackage[utf8x]{inputenc}
\usepackage{lineno}
\usepackage[T1]{fontenc}
\usepackage{mathptmx}
\usepackage[inline]{enumitem}

\begin{document}
\title{Emergent rhythms in coupled nonlinear oscillators due to dynamic interactions}

\author{Shiva Dixit}

\affiliation{Department of Physics, Central University of Rajasthan, NH-8,Bandar Sindri, Ajmer 305 817, India}

\author{Sayantan Nag Chowdhury}

\affiliation{Physics and Applied Mathematics Unit, Indian Statistical Institute, 203 B. T. Road, Kolkata-700108, India}

\author{Awadhesh Prasad}

\affiliation{Department of Physics and Astrophysics, University of Delhi, Delhi 110007, India}

\author{Dibakar Ghosh}
\email{dibakar@isical.ac.in}

\affiliation{Physics and Applied Mathematics Unit, Indian Statistical Institute, 203 B. T. Road, Kolkata-700108, India}

\author{Manish Dev Shrimali}
\email{shrimali@curaj.ac.in}

\affiliation{Department of Physics, Central University of Rajasthan, NH-8,Bandar Sindri, Ajmer 305 817, India}

\date{\today}

\begin{abstract}
	
The role of a new form of dynamic interaction is explored in a network of generic identical oscillators. The proposed design of dynamic coupling facilitates the onset of a plethora of asymptotic states including synchronous states, amplitude death states, oscillation death states, a mixed state (complete synchronized cluster and small amplitude desynchronized domain), and bistable states (coexistence of two attractors). The dynamical transitions from the oscillatory to death state are characterized using an average temporal interaction approximation, which agrees with the numerical results in temporal interaction. A first order phase transition behavior may change into a second order transition in spatial dynamic interaction solely depending on the choice of initial conditions in the bistable regime. However, this possible abrupt first order like transition is completely non-existent in the case of temporal dynamic interaction. Besides the study on periodic Stuart-Landau systems, we present results for paradigmatic chaotic model of  R\"{o}ssler oscillators and  Mac-arthur ecological model.

\end{abstract}

\maketitle

	\begin{quotation}
	{\bf  Population biology of ecological networks, person to person communication networks, brain functional networks, possibility of outbreaks and spreading of disease through human contact networks, to name but a few examples which attest to the importance of researches based on temporal interaction approach. Studies based on representating several complex systems as time-varying networks of dynamical units have been shown to be extremely beneficial in understanding real life processes. Surprisingly, in all the previous studies on time-varying interaction, death state receives little attention in a network of coupled oscillators. In addition, only a few studies on dynamic interaction have considered the proximity of the individual systems' trajectories in the context of their interaction. In this paper, we propose a simple yet effective dynamic interaction scheme among nonlinear oscillators, which is capable of relaxing the collective oscillatory dynamics towards the dynamical equilibrium under appropriate choices of parameters. The dynamics of coupled oscillators can show fascinating complex behaviors including various dynamical phenomena. A qualitative explanation of the numerical observation is validated through linear stability analysis and interestingly, a linear stability analysis is persued even when the system is time-dependent. An elaborate study is contemplated to reveal the influences of our proposed dynamic interaction in terms of all the network parameters.}
	
\end{quotation}

\section{Introduction} \label{introduction}

\par Time-varying interaction~\cite{holme2012temporal} enjoys a widespread recognition among researchers due to its numerous practical applications. Various interdisciplinary research approaches \cite{holme2012temporal,nag2020cooperation}, 
 from both theoretical and experimental points of view, offer fresh new insights about the collective phenomena due to time-varying interaction. Recently, dynamical systems are found to be an efficient and prominent tool, which open the door to study the role of dynamic interactions in a broad variety of complex systems. The interactions among dynamical systems can give rise to fascinating collective behavior ranging from synchronization~\cite{pikovsky2003synchronization,arenas2008synchronization,hramov2004approach}, extreme events~\cite{chowdhury2019extreme,9170822}, chimera states~\cite{ achimera, maksimenko2016excitation,khaleghi2019chimera,andreev2020stimulus,majhi2019chimera}, 
  suppression of oscillations~\cite{ad_report,chowdhury2020effect, od_report} to revival of oscillations \cite{zou2013reviving,zou2015restoration} and many more. Interestingly, most of the previous investigations among dynamical units are confined within the regime of static interactions. Contrary to this, in the present article, we bring the notion of dynamic interaction on collective behavior of coupled nonlinear dynamical systems.

\par The relevance of dynamic interaction has been recognized already by considering few general frameworks on the interacting nonlinear oscillators, where either the interacting function is changing over time~\cite{aprasad_pramana, yadav2017dynamics, sudhashu2019,sd1}, or the interaction depends on the states of the individual oscillators \cite{chowdhury2019convergence,schrodertransient, threshold_2019}. In this article, we consider a new form of dynamic mean-field interaction with two distinct possible variations. One of these implemented coupling configuration is that individual oscillators are interacting with mean-field coupling form \cite{winfree1967biological,mirollo1990amplitude} for a pre-specified certain time period and they remain isolated for the remaining time window. Another possibility is to introduce the dynamic interaction through the scenario, where individual oscillators are interacting only when the mean state of the oscillators lies within a certain vicinity of the phase space. This type of modulated interaction is quite common in robotic communication \cite{buscarino2006dynamical} as well as in wireless communication systems, where transmission is only activated within a particular region of the physical space or for a particular specified time \cite{chowdhury2019extreme, chowdhury2019convergence,schrodertransient, threshold_2019}. Instead of static (time-independent) coupling formalism, most of the realistic systems including physical, biological and social networks possess time-varying connectivity. Our imposed restricted interaction produces unanticipated dynamical states, that could not be expected if the interaction among those oscillators is possible in the entire time-domain, or in the whole state space. 
 In fact, there are some real instances, where it is not possible to have a continuous interaction for all the time and in the entire state-space due to the practical limitations \cite{tandon2016synchronizing}.  

\par Motivated by these facts, we try to capture the essence of realistic cases through the paradigm of dynamic interaction in coupled nonlinear oscillators. The remaining part of this paper is organized as follows. In Sec.\ \ref{model}, we discuss the proposed mechanism of dynamic mean-field interaction in the coupled nonlinear oscillators in detail, where the oscillator’s motion affects the network topology. This is followed by the detailed numerical investigations that are carried out 
for several systems including Stuart–Landau system \cite{kuramoto2003chemical} (limit-cycle oscillator), R\"ossler system \cite{rossler1976equation} (chaotic oscillator) and Mac-Arthur system \cite{goldwyn2008can, rosenzweig1963graphical} (ecological system).
The generic transitions from the oscillatory to steady state are also validated using linear stability analysis in Sec.\ ~\ref{NR}. Lastly, we conclude and summarize our results in Sec.\ ~\ref{conc}.

\section{Mathematical model} \label{model}


\par The time-evolution of each $i$-th oscillator $(i=1,2,\cdots,N)$ can be described by the following set of equations \cite{mirollo1990amplitude,sharma2012amplitude},
\begin{equation}
\begin{split}
\dot{\mathbf{X}}_{i} &={F}(\mathbf{X}_{i})+ \epsilon \beta ({H} \overline{\mathbf{X}} - \mathbf{X}_{i}),
\end{split}
\label{eq1}
\end{equation}
where $\mathbf{X}_i$ represents state variables of the $m$-dimensional $i$-th oscillator and $F: \mathbb{R}^m \to \mathbb{R}^m$ reflects the intrinsic dynamics of each node of the network \eqref{eq1}. The term $\overline{\mathbf{X}}= \frac{1}{N} \sum_{j=1}^{N} \mathbf{X}_j$ gives the arithmetic mean of the state variables and $N (\ge 2)$ is the number of independent non-linear dynamical systems. The control parameter $\epsilon$, reckoning as the strength of the interaction among those oscillators, is taken to be identical for all oscillators. $\beta$ is an $m \times m$ diagonal matrix with diagonal entries  
\[
\begin{cases}
\beta_{kk}=1, \hspace{0.5cm} \text{if the $k$-th component of the oscillator} \\ \hspace{1.9cm} \text{takes part in the mean-field coupling}\\
\beta_{kk}=0, \hspace{0.5cm} \text{otherwise.}
\end{cases}
\]
\par Our proposed dynamic interaction is determined by the step function $H$, which is considered to be a function of mean-field term $\overline{\mathbf{X}}$ and time $t$. The range of $H$ consists only two values $0$ and $1$. This function $H$ helps to treat our dynamic interaction policy as an on–off type of dynamic interaction \cite{schrodertransient}, where $H=0$ depicts the oscillators are completely independent of each other. While $H=1$ helps to sustain the global interaction with the mean field coupling. Thus, treating each oscillator of the coupled dynamical network \eqref{eq1} as a node \cite{threshold_2019}, the degree of each node is either $0$ (when $H=0$) or $N-1$ (when $H=1$). In what follows, we now introduce the step function $H$ in two different ways.

\subsection{Spatial Dynamic Interaction (SDI)} The state-space dependent interaction function $H=H(\overline{\mathbf{X}}, t)$ is defined as
\begin{eqnarray}
	H(\overline{\mathbf{X}}, t)=
\begin{cases}
1,  \text{if} \ \overline{\mathbf{X}} \in R' \\
0,  \text{if} \ \overline{\mathbf{X}} \notin R'
\end{cases}
\label{spatialeq3}
\end{eqnarray}
\noindent where $R'\subseteq \mathbb{R}^m$ is a subset of the state-space $\mathbb{R}^m$, where 
interaction is active. Here, the subset $R'$ can be defined in term of $\Delta^\prime$, which is 
written in the normalized form as $\Delta = \Delta^\prime/\Delta_a$, where $\Delta_a$ 
is the width of the attractor along the clipping direction~\cite{schrodertransient}. 

\par Control parameter $\Delta$ plays a decisive role by turning on the interaction among the dynamical units whenever they are inside a pre-specified subspace $R'$ of the phase space. Specifically, where time-independent and time-varying diffusive  interactions do not lead to stabilize the unstable stationary point of the uncoupled system \cite{chowdhury2019extreme,9170822,mirollo1990amplitude} in general, clipping in an interval through mean-field diffusive coupling is found to be beneficial, which can stabilize the unstable stationary points of the isolated systems. Here, for $\Delta$ $\to$ $0+$, there is no sufficient interaction between the units and as a result of our proposed network \eqref{eq1}, only self-negative feedback is activated. While for $\Delta \to 1-$, all oscillators are globally coupled with mean-field interaction.

\subsection{Temporal Dynamic Interaction (TDI)} On the other hand, we also consider the time-varying function $H=H(\overline{\mathbf{X}}, t)$ as a periodic step function of period $T$, which is defined as
\begin{eqnarray}
	H(\overline{\mathbf{X}}, t)=
\begin{cases}
1, \ \ \ \text{if} \ 0 < t \leq \tau^\prime \\
0, \ \ \ \text{if} \ \tau^\prime < t \leq T \\
\end{cases}
\label{temporaleq3}
\end{eqnarray}

\noindent where, $\tau=\tau^\prime/T \in [0, 1]$, $\tau^\prime$ is an active interaction time window
and $T (=2\pi/\omega)$ is an average time period of the oscillations of uncoupled system \cite{sd1, sd2}. 
Here, interaction is active for $\tau^\prime$ period of time, while it is inactive during 
the time $T - \tau^\prime$ as defined in Eq. (\ref{temporaleq3}).

\par Here, $\tau$ is an active interaction time-period, where the interaction switches periodically between the mean-field diffusive interaction and self-negative feedback. If the time period of the 
system is $T$, then the mean-field interaction is activated for fraction $\tau$ of the cycle, which is followed by self 
negative feedback for the remaining time window $(T-\tau)$. So when $\tau=0$, the oscillators are always under the effect 
of the negative self-feedback and when $\tau=1$,  the oscillators are always coupled through mean-field diffusive 
interaction for all the time. For $0<\tau<1$, the oscillators experience interaction through mean-field diffusive 
interaction for time $\tau T$ and negative self-feedback for time $T(1-\tau)$ in each cycle of period $T$. 
\par Hence, $\Delta$ and $\tau$ are two crucial control parameters 
of our model. In the following sections, our key interest will be to identify the emergent collective phenomena due to the interplay of different parameters $\Delta$, $\tau$ and the coupling strength $\epsilon$ for fixed $N$ number of non-linear oscillators. All these parameters play an important role in order to obtain different dynamical states. The numerical simulations are done using the Runge-Kutta fourth-order (RK4) method for a time of $10^5$ units with a fixed integration time $dt=0.01$ after removing enough transients ($\sim 10^6$ units).
\section{Results for Limit Cycle System: Coupled Stuart-Landau Oscillator} \label{NR}
\par We consider $N$ identical Stuart Landau (SL) oscillators $z_j$, $j=1,2,\cdots,N$ coupled through the spatio-temporal dynamic interaction. The dynamical equations of the coupled system are given as
\begin{equation}
\begin{split}
\dot{z}_{j} &= (\rho+i\omega-{|z_{j}|}^2)z_j+ \epsilon(H\overline{z}-Re(z_j)).  
\end{split}
\label{slmodel}
\end{equation}
Here, $j=1,2,\cdots,N$, where $N$ is the total number of oscillators in the dynamical network. $z_j= x_j+iy_j \in \mathbb{C}$ is the state variable of the $j$-th oscillator with $i=\sqrt{-1}$. $\overline{z}=\frac{1}{N} \sum_{k=1}^{N} Re(z_k) = \frac{1}{N} \sum_{k=1}^{N} x_k$ is the mean field term of the coupled system. The parameter $\rho \in \mathbb{R}$ acts like a bifurcation parameter and a Hopf bifurcation occurs at $\rho=0$. For $\rho \le 0$, each isolated individual $j$-th oscillator stabilizes into the trivial stationary point $z_j=0$, and for $\rho > 0$, single SL oscillator exhibits a stable periodic attractor with radius $\sqrt{\rho}$ and eigen frequency $\omega$ \cite{kuramoto2003chemical}. The function $H$ is considered as a function of $Re(z_k)$, $(k=1,2,\cdots,N)$ and time $t$. We investigate the collective behavior of the coupled SL oscillators with dynamic spatial and temporal interaction by varying control parameters such as $\rho$, $\omega$, $\Delta$, $\tau$ and $\epsilon$. For the numerical simulations, initial conditions are chosen randomly from the interval $[-1,1] \times [-1,1]$. Without loss of any generality, $\rho=1$ and $\omega=2$ are taken for the following numerical investigations (unless stated otherwise).

\begin{figure}[!t]
	\centerline{\includegraphics[width=0.35\textwidth]{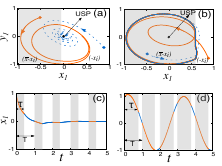}}
	\caption{{\bf Transient trajectories}: The dynamics of the first oscillator of the coupled SL oscillator is shown here without loss of any generality. The interaction is activated in the shaded region (gray) \textbf{(a)} for $\Delta=0.4$, and \textbf{(b)} for $\Delta=0.8$ . The solid orange curve is for that trajectory, which start initially at the coupling activated region (gray shaded region). Whereas, the dashed blue curve represents a trajectory whose initial condition lie outside of the gray shaded region. Here, USP stands for unstable stationary point of the isolated SL oscillator. Here, the subfigures are for \textbf{(c)} $\tau=0.4$, and \textbf{(d)} $\tau=0.8$, where the interaction is activated in the $\tau$ period of time while absent for the rest of the time-period. For all the subfigures, $N=100$ and $\epsilon=2.1$.}
	\label{visualsp}
\end{figure}

\subsection{Spatial dynamic interaction (SDI): } \label{SDI}
\par To illustrate the effect of SDI, two distinct Figs.\ \ref{visualsp} (a) and \ref{visualsp} (b) are presented for fixed coupling strength $\epsilon=2.1$ with $\Delta=0.4$ and $\Delta=0.8$, respectively. Here, minima and maxima of the attractor (with $\rho=1$) are  $-1$ and $1$, respectively. Then, $\Delta_a=2$ and thus, $\Delta$ assigns the active region $[\text{global minima of the attractor}, \text{ global minima of the attractor}+\Delta_a \times \Delta]$ as per our implementation. Hence, $\Delta=0.4$ implies that the coupling activated sub-space $(H=1)$ is $[-1, -0.2]$. Similarly, $\Delta=0.8$ means that coupling is turned on only within the interval $[-1, 0.6]$. Clearly, the dynamic coupling is active only within the gray-shaded region and it is turned off $(H=0)$ outside of that region in Figs.\ \ref{visualsp} (a) and \ref{visualsp} (b). Note that, a key difference is observed between $\Delta=0.4$ and $\Delta=0.8$. For $\Delta=0.4$, the coupling-activated subspace does not contain the origin, the unstable stationary point of the isolated SL oscillator. For this specific choice of $\Delta=0.4$, the system \eqref{slmodel} displays a bistable behavior. To demonstrate this feature, we choose two different initial values of $(x_1,y_1)$. One is from the interaction active region $[-1, -0.2]$ and the other one is from $(-0.2, 1]$. The initial choice of $(x_1,y_1)$ from the coupling active region gives rise to a limit cycle (solid orange), whereas the initial condition for $(x_1,y_1)$ from the inactive region leads to cessation of oscillations (dashed blue) under the influence of only negative self-feedback and the system consequently converges to the origin. This bistable behavior is completely vanished for $\Delta=0.8$. For $\Delta=0.8$, the trajectories always settle down to sinusoidal like oscillations irrespective of the choice of initial conditions belonging to $[-1,1] \times [-1,1]$.

\par
\begin{figure}[!t]
	\centerline{\includegraphics[width=0.4\textwidth]{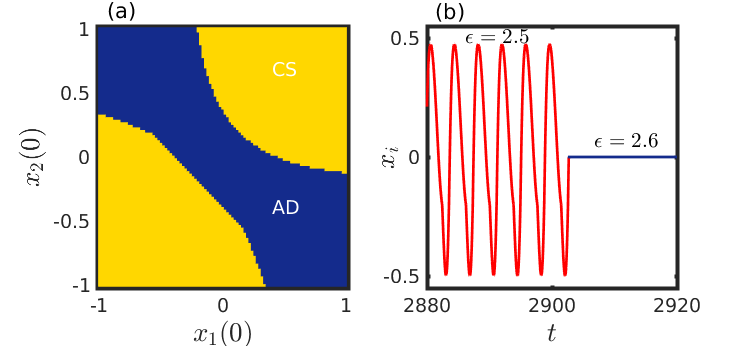}}
	\caption{ (a) {\bf Effect of initial conditions}: Here, $\Delta=0.4$ and $\epsilon=2.1$ are set as taken in Fig.\ \ref{visualsp} (a). For better visualization, only $N=2$ oscillators are taken to plot this basin of attraction. $y_1(0)=0.1$ and $y_2(0)=0.0$ are kept fixed. $x_i(0)$ are varied uniformly from $-1$ to $1$ with fixed step-length $0.02$ for $i=1,2$. For $\overline{z}(0) = \dfrac{x_1(0)+x_2(0)}{2} \in [-0.33,0.43]$ (approximately), the only stable attractor is the amplitude death (AD) state (blue). Beyond this region, synchronized limit cycles (CS) are found. (b) {\bf Time–series near the transition point for two coupled SL oscillators}: Here, the initial condition is $(x_1(0),y_1(0),x_2(0),y_2(0))=(-1.0,0.1,-1.0,0.0)$ and $\Delta=0.4$. With slight enhancement of coupling strength $\epsilon$ from $2.5$ to $2.6$, both the oscillators settle down to a common stable death state from the synchronized oscillatory behavior. }
	\label{basin}
\end{figure}

 To explore the effect of initial conditions, we plot the basin of attraction with respect to the variables $x_1$-$x_2$ in Fig.\ \ref{basin} (a) by keeping fixed the values of the other variables $y_1(0)=0.1$ and $y_2(0)=0.0$. For simplicity, only $N=2$ oscillators are considered. Although, it seems in Fig.\ \ref{visualsp} (a) for these values of $\Delta=0.4$ and $\epsilon=2.1$, the choice of $(x_1(0),y_1(0))$ from the inactive region $(-0.2,1.0]$ always leads to AD, but that is not necessarily reflect the true story. Definitely, the stabilization of AD state depends not only on the initial condition of any single oscillator. Other variable’s initial conditions as well as the velocity fields near the interaction switching on-off region are also equally important to stabilize the death state. For the particular choice of $y_1(0)=0.1$ and $y_2(0)=0.0$, the basin of the coexisting attractors reveals that suppression of oscillations to AD is possible if $\overline{z}(0) = \dfrac{x_1(0)+x_2(0)}{2} \in [-0.33,0.43]$. However, this bistable behavior is completely suppressed with increasing coupling strength $\epsilon$. An example is depicted in Fig.\ \ref{basin} (b) with $N=2$ oscillators and fixed initial conditions. A sudden dynamical transition is portrayed, where the system exhibits oscillatory state for $\epsilon=2.5$, but the oscillatory solutions lose their stability for $\epsilon=2.6$ and homogeneous steady state is found.
\begin{figure}[!t]
	\centerline{\includegraphics[width=0.3\textwidth]{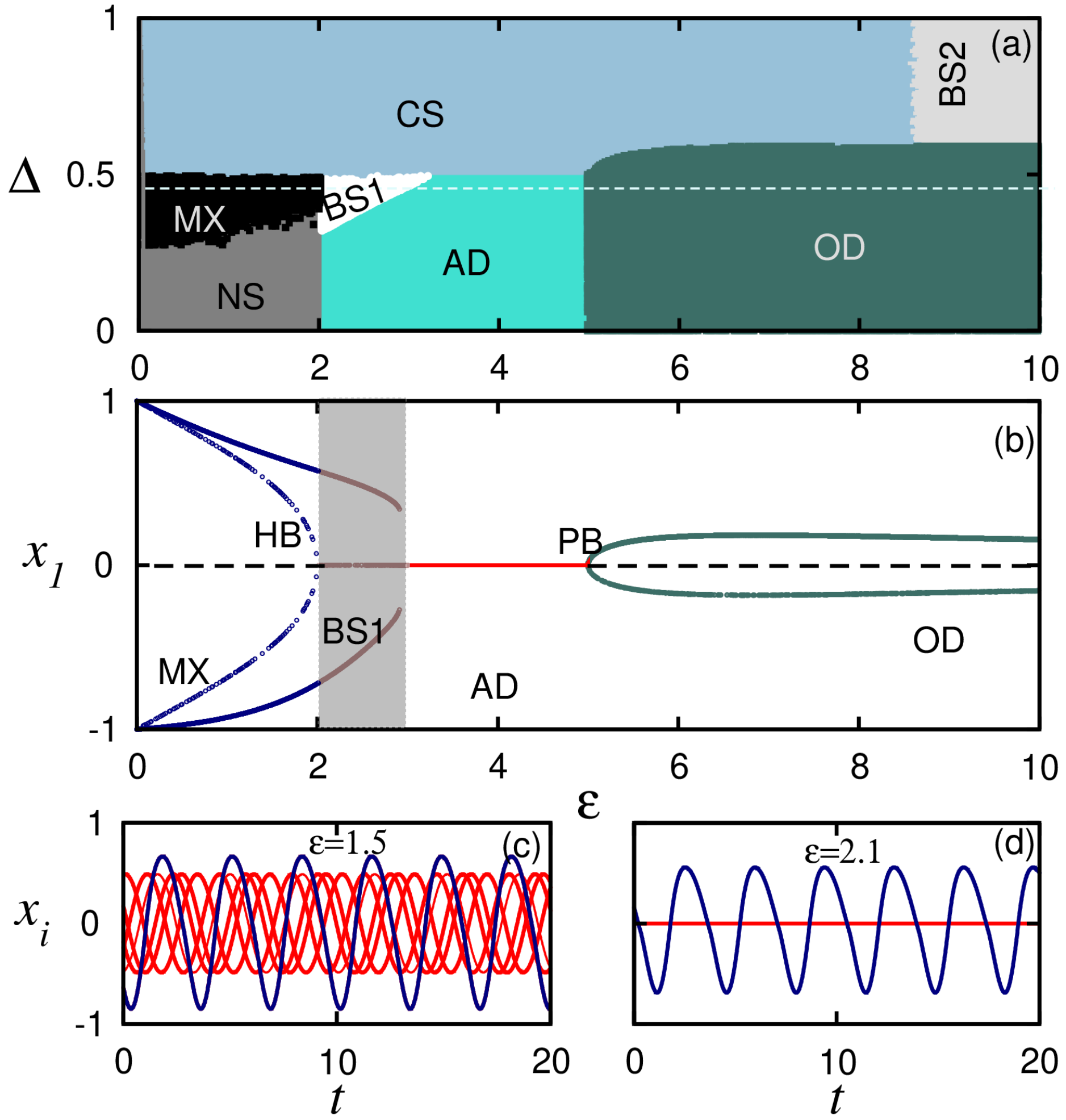}}
	\caption{ (a) {\bf Diverse emergent dynamical behaviors of $N=100$ coupled SL oscillators in the parameter plane $\mathbf{\epsilon}$-${\Delta}$}: Numerical simulation of the asymptotic behavior of the system \eqref{slmodel} gives rise to several dynamical states including NS (desynchronized oscillatory state), CS (complete synchronized oscillatory state), AD (amplitude death state), OD (oscillation death state), MX (mixed state, where desynchronized small oscillations and  a group of completely synchronized large oscillations co-exist), BS1 (bistable regime with complete synchronized state and amplitude death state) and BS2 (where complete synchronized state and oscillation death state co-exist).  (b) {\bf Bifurcation diagram of the system \eqref{eq1} with respect to the coupling strength $\epsilon$ for $\Delta=0.45$}: Shaded region is the BS1 state. Here, solid line corresponds to the stable behavior and dashed line represents unstability of the origin. HB and PB stand for Hopf bifurcation and pitchfork bifurcation, respectively. (c)-(d) {\bf Temporal evolution of $x_i$ for $i=1,2,\cdots,N$}: Time traces are shown in (c) for $\epsilon=1.5$ and  in (d) for $\epsilon=2.1$. Complete synchronization (blue) or partial synchronization (red) is observed in the subfigure (c) depending on the suitable initial conditions. Similarly, depending upon the initial states, completely synchronized limit cycle (blue) or AD state (red) is appeared in the subfigure (d). The other parameters are $\rho=1, \omega=2$, and $N=100$. For details see the text.}
	\label{slepsvsdel}
\end{figure}
\par To understand the effect of the SDI in $N=100$ coupled SL oscillators, we analyze the interplay of the parameters  $\epsilon$ and $\Delta$ in Fig.\ \ref{slepsvsdel} (a). We find here that there exists a critical value $\Delta^* \approx 0.49$ that designates two major transition scenarios. For $\Delta>0.49$, the coupled SL oscillators are oscillating coherently at very small non-zero coupling strength $\epsilon$. With increment of $\epsilon$, a transition takes place from the synchronized state (CS) \cite{pikovsky2003synchronization} to oscillation death (OD) \cite{koseska2013oscillation} for $\Delta>0.49$. Even after a certain threshold of $\epsilon$, the coexistence of oscillatory states and OD states are observed in the parameter plane $\epsilon-\Delta$, which is shown as $BS2$ region in Fig.\ \ref{slepsvsdel} (a). For $\Delta<0.49$, the suppression of oscillation from the oscillatory state is also perceived for $\epsilon \ge 2.0$. The Jacobian matrix of the coupled systems \eqref{slmodel} with $H=0$ at the trivial stationary point $( \underbrace{{\bf O},{\bf O},\cdots,{\bf O}}_\text{N times})$, where ${\bf O}=(0,0) $  is the unstable stationary point of the SL oscillator, is given by the block diagonal matrix $A \oplus A \oplus A \cdots \oplus A$ ($N$ times). Here, $A$ is the Jacobian matrix of the isolated system with only negative self-feedback at ${\bf O}=(0,0) $. The eigen values of $A$ are 
\begin{eqnarray}
	\lambda_{1,2} =  \rho - \dfrac{\epsilon}{2} \pm \dfrac{1}{2} \sqrt{(\epsilon-2\omega)(\epsilon+2\omega)}.
\label{eq9}
\end{eqnarray}
This eigenvalue analysis suggests that amplitude death (AD) \cite{ad_report,karnatak2007amplitude} is impossible for $\omega \le 1$. For any value of $\omega > 1$, equating the real part of $\lambda_{1,2}$ to zero, we get $\epsilon_{HB} = 2\rho$. Setting $\lambda_{1,2}=0$, we find $\epsilon_{PB}=\rho + \dfrac{\omega^2}{\rho}$, where $\rho \ne 0$. These bifurcation points $\epsilon_{HB} = 2$ and $\epsilon_{PB}=5$ for $\rho=1$ and $\omega=2$ fit exactly with our numerically obtained bifurcation diagram given in Fig.~\ref{slepsvsdel} (b). The dashed line in Fig.~\ref{slepsvsdel} (a) corresponds to the choice of $\Delta=0.45$ at which the bifurcation diagram (Fig.~\ref{slepsvsdel} (b)) is scrutinized. An inverse Hopf bifurcation occurs at $\epsilon_{HB} = 2$ and the stable periodic attractor is suppressed and give birth to a stable homogeneous steady state (AD). Depending on the coupling parameter, a symmetry breaking pitchfork bifurcation occurs at $\epsilon_{PB}=5$ and as a result of that, coupling dependent inhomogeneous steady states emerge. This transition from AD to OD is also noticed in Fig.~\ref{slepsvsdel} (a) for $\Delta < 0.49$. It should be noted that for $\Delta < 0.49$, the coupling activated region does not contain the unstable stationary point ${\bf O}$, and thus it is highly probable that the only coupling remains active for $\Delta < 0.49$ is negative self-feedback. However, a bi-stable region $BS1$ is recognized in Fig.~\ref{slepsvsdel} (a), where the limit cycle and AD states can coexist. This bistable region (shaded region) is also found in Fig.~\ref{slepsvsdel} (b). To represent our findings, the temporal evolution of the variable $x_i$ is shown in Figs.\ \ref{slepsvsdel} (c) and \ref{slepsvsdel} (d) at	$\epsilon=1.5$ and $\epsilon=2.1$ for $\Delta=0.45$, respectively. Figure \ref{slepsvsdel} (c) depicts incoherent nature of the trajectories (red curves), where each SL oscillator exhibits stable periodic orbit, but of various radii (amplitudes). Depending on the suitable initial conditions, $N=100$ trajectories may collapsed into a single trajectory (blue curve) as shown in Fig. \ref{slepsvsdel} (c). This region of mixed state (MX) is highlighted in Fig.~\ref{slepsvsdel} (a). The resulting time series for $\epsilon = 2.1$ in Fig.~\ref{slepsvsdel} (d) reflects the coexistence of synchronized state and AD state, which is the signature of $BS1$ region.
\subsection{Temporal dynamic interaction (TDI):}  \label{TDI}
\par The continuous interaction is not always existent and
manageable in many real systems, such as the transmissions of biological signals between synapses and
the communications of ant colonies in the processing of migration, as well as the seasonal interactions
between predator–prey in the ecosystem, leading to the discontinuous and intermittent coupling relationship \cite{sun2018inducing}.  Therefore,
it is of essential importance to investigate the oscillation patterns in the coupled system containing temporal discontinuous
coupling. Figures \ref{visualsp} (c) and \ref{visualsp} (d) represent the effect of our considered time-varying interaction through TDI. Figure \ref{visualsp} (c) exhibits that the interaction remains constant in
one part of the period where $\tau=0.4$, and for the remaining part $(1-\tau)=0.6$, the interaction disappears. In Fig.\ \ref{visualsp} (d), the same process is repeated for $\tau=0.8$. Here, we have taken $T=1.0$ for easier demonstration.
\begin{figure}[!t]
	\centerline{\includegraphics[width=0.3\textwidth]{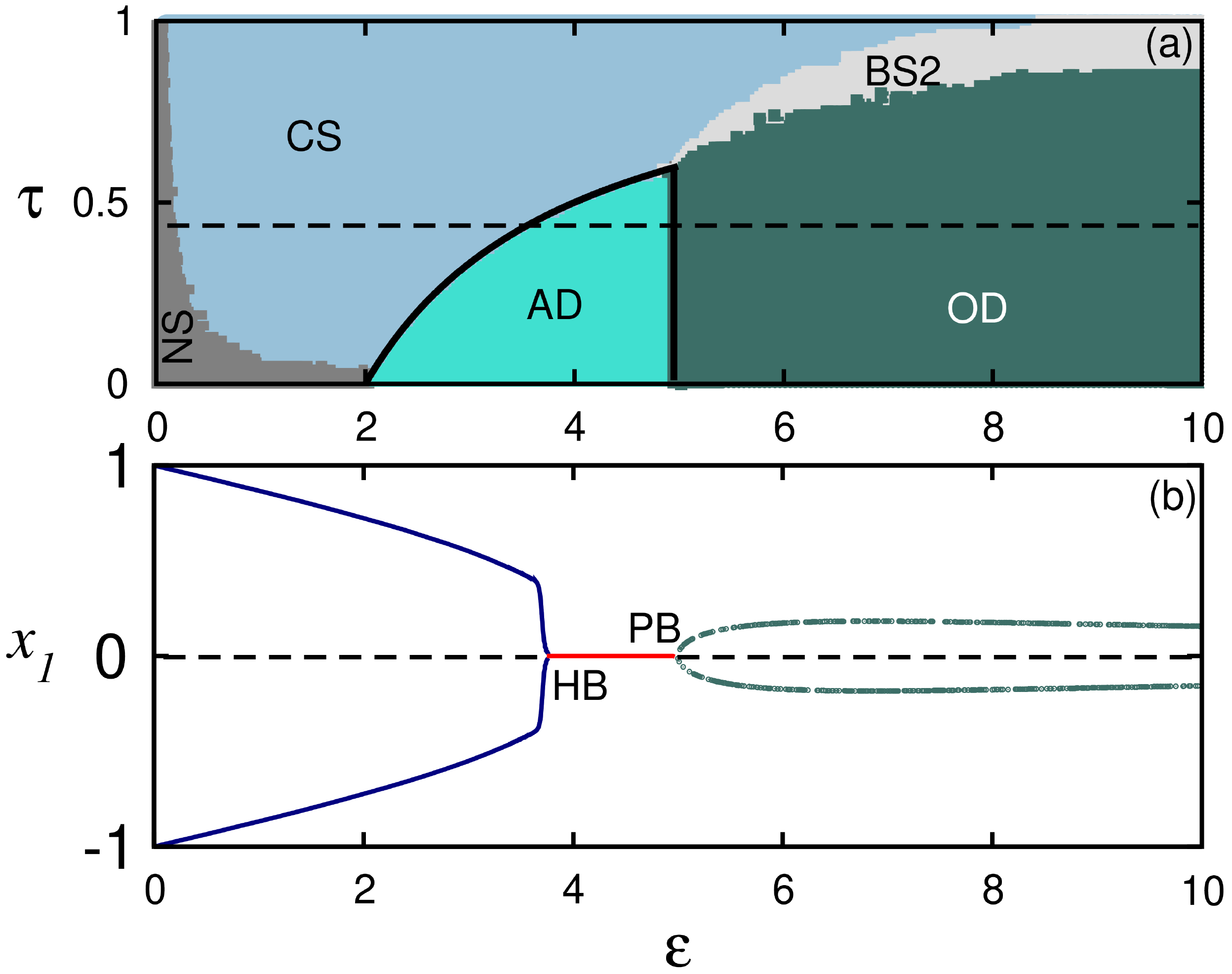}}
	\caption{  (a) {\bf Map of dynamic regimes for $N=100$ coupled SL oscillators in the parameter plane $\tau-\epsilon$}: The regimes marked as NS, CS, AD, OD and BS2 representing the desynchronized oscillatory state, complete synchronized oscillatory state, amplitude death state, oscillation death state, and bistable state (oscillatory state and oscillation death state), respectively. The solid black line is the analytically derived relation \eqref{condtemp}. (b) {\bf Bifurcation diagram of $N=100$ coupled SL oscillators with respect to coupling strength $\epsilon$ for $\tau=0.45$}: HB and PB are Hopf bifurcation and pitchfork bifurcation point, respectively. The coupled system is first stabilized at the origin through inverse Hopf bifurcation and subsequently, OD states are born through a pitchfork bifurcation. For details, please see the text.}
	\label{epsvstau}
\end{figure}
To illustrate the effect of TDI, $H=$ $H(t)$ is taken into consideration as defined in Eq.~(\ref{temporaleq3}). Here, $H=$ $H(t)$ depends on the interaction active time $\tau$ and time period $T$ of the system on the network \eqref{eq1} of SL oscillators. Here, we consider smaller values of $T$ as compared to the oscillation time period ($T_{SL} \sim 1.26$) of uncoupled oscillator~\cite{sd1}. The results are shown here for $T=0.1$. We obtain the parameter plane between $\epsilon-\tau$ for coupled SL oscillator in Fig.~\ref{epsvstau} (a).  A transition is witnessed from incoherent state (NS) to synchronized state (CS) with increasing coupling strength $\epsilon$. With further increment of $\epsilon$, either OD states or coexistence of oscillatory state and OD state is found depending on the value of $\tau$. The bistable region is marked as BS2 in Fig.~\ref{epsvstau} (a). To further understand the scenario, $\tau=0.45$ is chosen along the black dashed line in Fig.~\ref{epsvstau} (a) and the bifurcation diagram is plotted at this value of $\tau$ in Fig.~\ref{epsvstau} (b). Increment of $\epsilon$ reveals the suppression of stable limit cycle and AD appears at $\epsilon=2$ for $\tau=0$, through inverse Hopf bifurcation as shown in Fig.~\ref{epsvstau} (a). While at $\epsilon = 5$ through a super-critical pitchfork bifurcation, origin becomes unstable and two new stable states are created, giving birth to OD. 

\par Usually, a linear stability analysis is carried out at a stationary point of the system which is time-independent. In this case as the interaction (i.e., \ $h(t)$) in Eq.~(\ref{slmodel}) is time dependent, we consider an average eigenvalue $\lambda = \frac{[\tau^{'} \lambda_{on}+(T-\tau^{'})\lambda_{off}]}{T}$, 
where $\lambda_{on}$ and $\lambda_{off}$ are the numerically largest eigenvalues of the stability matrix at the stationary point over the period $\tau^{'}$ and $T-\tau^{'}$, respectively~\cite{aprasad_pramana}. Therefore, the linear stability analysis at a stationary point (zero in Eq.\ \eqref{slmodel}) provides nontrivial characteristic equations,
\begin{eqnarray}
	\lambda =  \rho \tau  + (1-\tau) \Bigg(\frac{2 \rho - \epsilon}{2}\Bigg). 
\label{cond}
\end{eqnarray}
Letting $\lambda = \alpha + i \gamma$, where $\alpha$ and $\gamma$ are real and imaginary part of the eigenvalues, Eq.~(\ref{cond}) leads to 
\begin{eqnarray}
	\alpha = 2 \rho - \epsilon + \epsilon \tau.
\label{condf}
\end{eqnarray}
The solid black line in Fig.~\ref{epsvstau}, corresponds to the locus $(\alpha = 0$ in  Eq.~(\ref{condf}))
\begin{eqnarray}
	\tau = 1-\frac{2 \rho}{\epsilon}, \hspace{0.2cm} \epsilon \ne 0.
\label{condtemp}
\end{eqnarray}
Analytical condition (solid black line) of Eq.~(\ref{condtemp}) matches perfectly with the numerically calculated amplitude death (AD) region in Fig.~\ref{epsvstau} (a). Here the dynamics changes from periodic attractor to AD via Hopf bifurcation as a real part of the eigenvalue $\alpha$ becomes negative. 

\subsection{Average interaction time in spatial dynamic interaction}
\par In addition, we have explored the relation between the average interaction time in SDI. It is obvious that whenever there is a discontinuous interaction in space, the temporal discontinuity must accompany it. In SDI, the interaction term is switched on or off depending on the mean state of the trajectories in the phase space, but in TDI, the on-off factor appears in a completely periodic manner. Here, we try to provide a correlation between the average interaction time in the spatial scheme ($\tau_{avg}(\Delta)$) and the active interaction time in temporal framework ($\tau$). 
\begin{figure}[!t]
	\centerline{\includegraphics[width=0.3\textwidth]{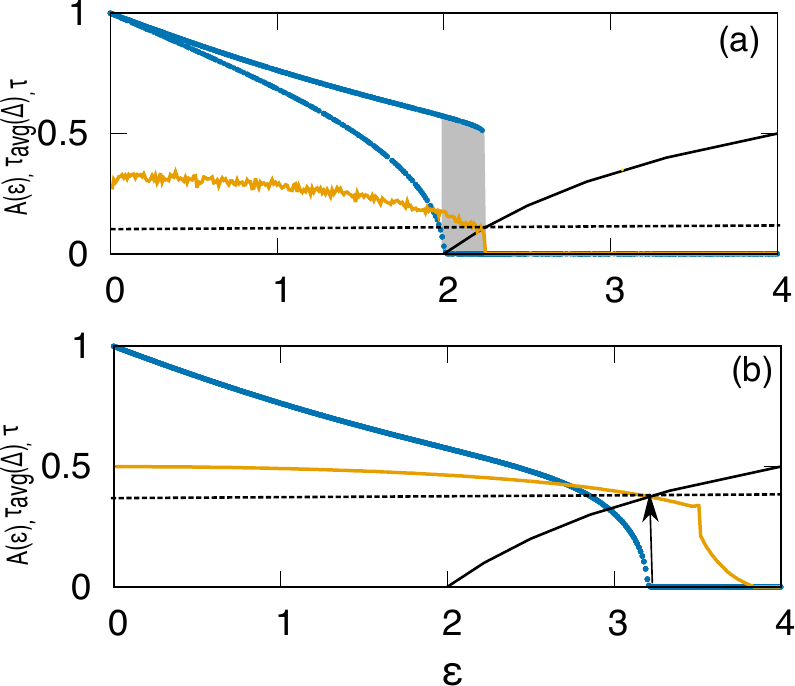}}
	\caption{ {\bf Average interaction time and average amplitude due to SDI}: The $\tau_{avg}(\Delta)$ (orange), $A$ (blue) and Eq.\ \eqref{condtemp} (black solid line) are plotted as a function of coupling strength $\epsilon$ at \textbf{(a)} $\Delta$ $= 0.35$ and \textbf{(b)} $\Delta$ $= 0.5$, respectively. Gray shaded region is the bistable region. Surprisingly, SDI yields both first-order and second-order transitions from oscillatory state to death state as observed in subfigure (a) depending on the initial conditions as well as the velocity fields near the interaction switching on-off region for $\Delta =0.35$. Although, that discontinuous and abrupt transition is completely vanished for $\Delta \ge 0.5$ as shown in subfigure (b) for $\Delta=0.5$. For details, please see the text. }
	\label{sptm}
\end{figure}
\par The order parameter $A$, which is the normalized average amplitude is now defined as
\begin{eqnarray}
	A=\frac{a(\epsilon)}{a(0)},
\label{contd}
\end{eqnarray}
where $a(\epsilon)=\dfrac{\sum_{i=1}^{N} (\langle x_{i,max}  \rangle_t-\langle x_{i,min}\rangle_t)}{N}$ \cite{sharma2012amplitude}. Here, $a(\epsilon)$ denotes the difference between the global maximum and minimum values of the attractor at a particular value of the coupling strength $\epsilon$ and $\langle \cdots \rangle_t$ indicates the sufficiently long time average. Thus, $A$ measures the average amplitude of the oscillators in the coupled system and for an oscillatory state, the value of $A$ will be greater than zero, while for a death state $A=0$ \cite{verma2017explosive}. We calculate the average on time $\tau_{avg}(\Delta)$ in space for a chosen region $(\Delta)$ over large number of initial conditions $(\approx 1000)$. In Fig.~\ref{sptm}, $\tau_{avg}(\Delta)$ (orange), average amplitude $A$ (blue) for a given $\Delta$ and the critical curve \eqref{condtemp} (black) separating the steady state and oscillatory regions in temporal interaction are plotted as a function of coupling strength $\epsilon$. As depicted in Fig.~\ref{sptm} (a), we find that at $\epsilon=0$, $\Delta = \tau_{avg}(\Delta)=0.35$. While with increasing $\epsilon$, the value of $\tau_{avg}(\Delta)$ eventually decreases and at a critical coupling strength $\epsilon_c(\Delta)$, the $\tau_{avg}(\Delta)$ reaches to $0$. The $\tau_{avg}(\Delta)$ crosses the analytical curve (Eq.\ \eqref{condtemp}) of the temporal interaction at $\epsilon \approxeq 2.23$, and after that, system completely ceases down to steady state. Beyond $\epsilon=2$, there is a gray shaded region of bistability in Fig.~\ref{sptm} (a), where both the stationary point attractor and limit cycle coexist. In this bistable region, where two behaviors exist side-by-side over a parameter region, a first order phase-transition to AD state is also uncovered in Fig.~\ref{sptm} (a) through an abrupt transition of $A$. The justification behind this discontinuous jump is due to the bistable behavior of the system as shown in Figs.\ \ref{basin} (a) and \ref{slepsvsdel} (a).  The traditional continuous transition is feasible based on the suitable choices of initial conditions as shown in Fig.~\ref{sptm} (a). But, there still exists a suitable set of initial conditions as shown in Figs.\ \ref{basin} (a) and \ref{slepsvsdel} (a), for which the system may still exhibit oscillation with small amplitude beyond the coupling strength $\epsilon_{HB}=2$. However, with increasing coupling strength $\epsilon$ beyond the critical value $\epsilon \approx 2.23$, the variation of $A(\epsilon)$ clearly indicates an abrupt transition from oscillatory state to death state. Such a sudden transition is also presented for two coupled SL oscillators with $\Delta=0.4$ in Fig.\ \ref{basin} (b). Coupled temporal system shows synchronized oscillatory behavior before the transition point at which all oscillators settle down to a common stable steady state. Thus, we see a second order transition or discontinuous transition from oscillatory state to death state completely depending on the initial conditions for these fixed value of $\Delta=0.35$. However, this interesting feature of first-order transition from oscillatory state to death state is completely lost for $\Delta \geq 0.5$. In Fig.~\ref{sptm} (b), one can see that $\Delta = \tau_{avg}(\Delta) = 0.5$ holds for a low coupling strength. But, as soon as $\epsilon$ increases, then the $\tau_{avg}(\Delta)$ gradually reduces and crosses the boundary condition (Eq.\ \eqref{condtemp}) of the temporal interaction at $\epsilon \approxeq 3.62$.  On the basis of the above analysis, in both the cases it is clear that whenever the $\tau_{avg}(\Delta)$ crosses the boundary (Eq.\ \eqref{condtemp}) of the temporal interaction, there is always a steady state arises in spatial interaction. This attests the well agreement of the numerical simulation with our analytically derived result (Eq.\ \eqref{condtemp}). 

\subsection{Effect of $\mathbf{\rho}$ and $\mathbf{\omega}$}

\begin{figure}[!t]
	\centerline{\includegraphics[width=0.4\textwidth]{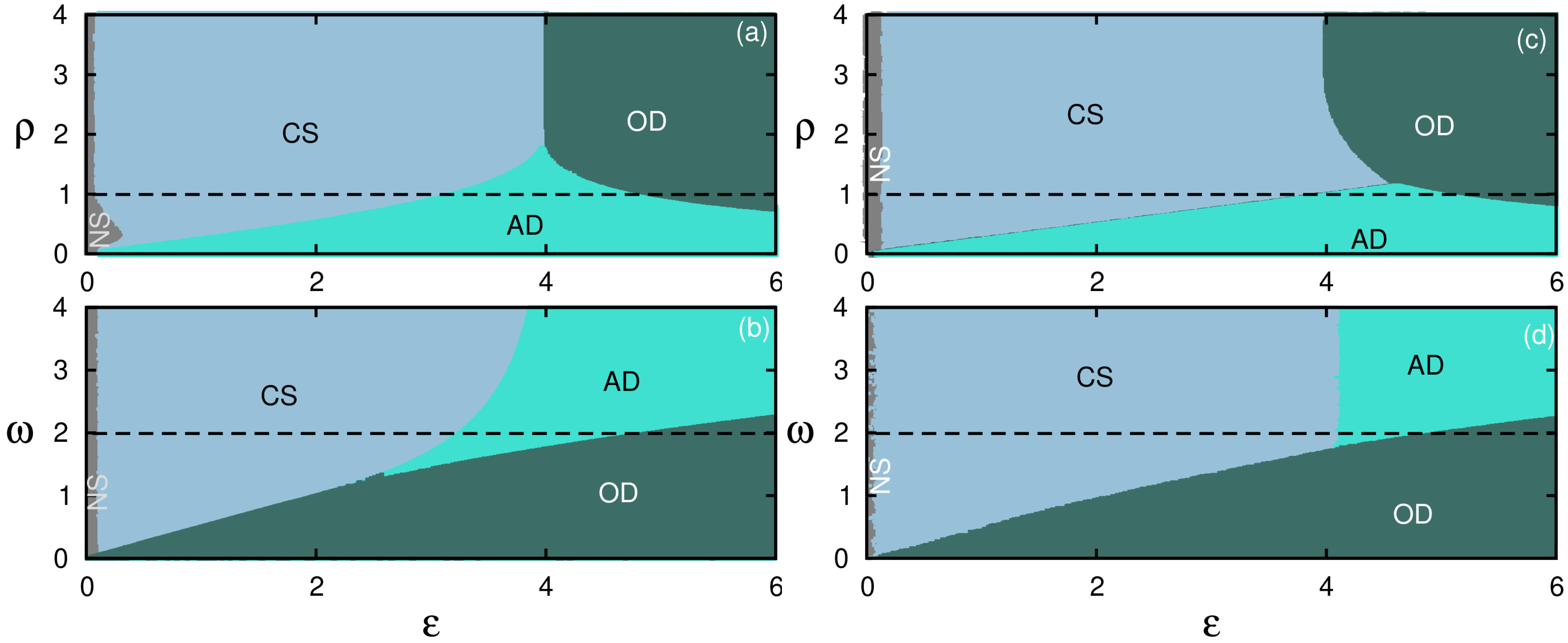}}
	\caption{ {\bf The impact of internal parameters $\rho$ and $\omega$ on the suppression of oscillations in an ensemble of SL oscillators}: The values of the parameters are taken as (a) $\Delta=0.5$ and $\omega=2.0$, (b) $\Delta=0.5$ and $\rho=1$, (c) $\tau=0.5$ and $\omega=2.0$, and (d) $\tau=0.5$ and $\rho=1$. Our proposed SDI and TDI are found to be robust over a large interval of internal parameters $\rho$ and $\omega$ in order to obtain the general transition from incoherent oscillatory state to stable death state. The dynamic coupling seems to break the inherent symmetry of the oscillator and thus gives rise to stable AD or OD states depending on the parameters. The regions NS and CS depict incoherent domain and coherent regime of synchronized limit cycle respectively.}
	\label{fig5}
\end{figure}	

\par Till now, the numerical results are presented with fixed internal parameters $\rho=1$ and $\omega=2$. These values are highlighted through black dashed lines in the subfigures of Fig.\ \ref{fig5}. Figure \ref{fig5} demonstrates the consequence of different choices of $\rho$ and $\omega$. The subfigures (a) and (b) of Fig.\ \ref{fig5} are drawn with fixed $\Delta=0.5$ and the remaining subfigures (c) and (d) depict the results for fixed $\tau=0.5$. All these subfigures portray the fact that the transition from the oscillatory dynamics to steady state is generic for all values of internal parameters. Although that steady states portray AD or OD depending on the values of the parameters. $\omega=2$ is kept fixed for Figs.\ \ref{fig5} (a) and (c) and $\rho=1$ is set for Figs.\ \ref{fig5} (b) and (d). For $\omega \le 1$, AD state is not found in Fig.\ \ref{fig5} (b), which agrees well with our eigenvalue analysis given in Sec.\ \ref{SDI}.
\begin{figure}[!t]
	\centerline{\includegraphics[width=0.3\textwidth]{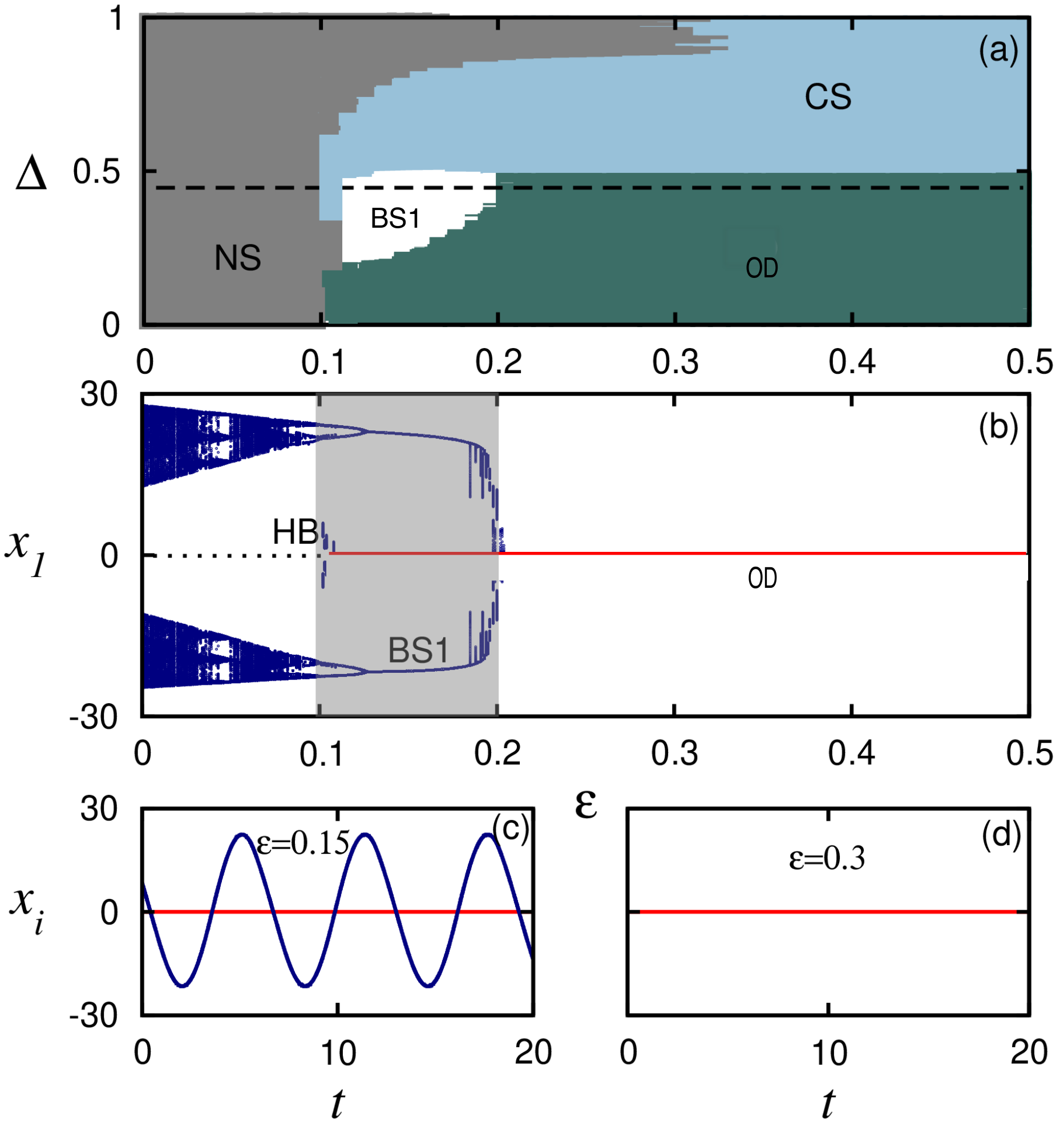}}
	\caption{  (a) {\bf Coupling strength $\epsilon$ vs SDI parameter $\Delta$ for identical R\"ossler oscillators with $N=100$}:  NS, CS, OD, and BS1 indicate desynchronized oscillatory state, complete synchronized oscillatory state, oscillation death state, and bi-stable state, respectively. (b) {\bf $x_1$ as a function of coupling strength $\epsilon$ with fixed $\Delta=0.45$}:	$x_1$ is plotted for stationary point solutions and extremum values for time dependent solutions of the coupled system \eqref{rosseq}. BS1 (the shaded region) shows the bistable regime, where complete synchronized state and oscillation death state may co-exist. (c)-(d) {\bf The time-series of variables $x_i$}: Different trajectories are converging to different attractors for different initial states in the subfigure (c). The coupling strength $\epsilon$ is $0.15$ for the subfigure (c) and $0.3$ for the subfigure (d), respectively. The system \eqref{eq1} may oscillate synchronously, or may converge to death state depending on appropriate initial conditions as revealed through the subfigure (c). Damped oscillation in the form of OD is observed in the subfigure (d). The other parameters are $a=0.1$, $b=0.1$ and $c=18.0$ and $N=100$.}
	\label{rosssp}
\end{figure}

\subsection{Chaotic System: Coupled R\"ossler Oscillator}

\par In order to further validate the generic nature of the transition from oscillatory state to death state, we examine the dynamic interaction on the coupled chaotic system. We consider $N=100$ coupled R\"ossler oscillators \cite{rossler1976equation}, interacting through spatial or temporal mean-field diffusive interaction. The dynamical equations are given as,
\begin{equation}
\begin{split}
\dot{x}_{i} &= -y_{i}-z_{i} + \epsilon(H\overline{x}-x_i), \\
\dot{y}_{i} &= x_{i}+a y_{i} ,        \\
\dot{z}_{i} &= b+z_{i}(x_{i}-c).       
\end{split}
\label{rosseq}
\end{equation}
\noindent The parameters $a=0.1, b=0.1$ and $c=18.0$ are set in the chaotic regime of isolated R\"ossler oscillator. The step function $H$ reflects the space and time dependent interaction as described earlier through the relations \eqref{spatialeq3} and \eqref{temporaleq3}, respectively. To illustrate the effect of space dependent interaction, we draw the phase-diagram $\epsilon-\Delta$ for $N=100$ coupled R\"ossler oscillators in Fig.~\ref{rosssp} (a). For $\Delta>0.49$, the system traverses from the desynchrony to the synchrony regime. But for $\Delta<0.49$, the chaotic oscillators can attain stable OD state at a suitable strength of $\epsilon$. This transition from oscillatory state to OD state can take place via first order or second order depending upon the value of $\Delta$ at lower coupling strength. The coexistence of the oscillatory and OD states are marked as $BS1$ in Fig.~\ref{rosssp} (a).

For $H=0$, the stationary point solutions of the Eq.\ \eqref{rosseq} are given by

$x^* = - \Bigg( \dfrac{P \pm \sqrt{ P^2-4abQ}}{2Q}\Bigg), 
y^* = -x^* / a$, and $z^* = -b/(x^*-c)$ with $P = c (a \epsilon - 1) $ and $ Q = - P/c$. The eigenvalues of the system at approximate value of stationary point for a given set of parameter values are, 
\begin{eqnarray}
	\lambda_{1} = -c,\;\;  
	\lambda_{2,3} &=& \frac{1}{2}\Big( a- \epsilon \pm \sqrt{(a + \epsilon - 2) (a + \epsilon + 2)  }   \Big)  
\label{eq91}
\end{eqnarray}
\par Thus equating the real parts of the complex eigen values of \eqref{eq91}, we obtain the condition of Hopf bifurcation as $\epsilon_{HB}=a=0.1$. This bifurcation point agrees quite well in Fig.~\ref{rosssp} (a). In fact, for $\Delta=0.45$ (the dashed line in Fig.~\ref{rosssp} (a)), we draw numerically the bifurcation diagram of coupled R\"ossler oscillators in Fig.~\ref{rosssp} (b). This bifurcation diagram of $x_1$ displays quenching of oscillation and gives birth to stable OD through inverse Hopf bifurcation. Note that, there exists a region of $\epsilon \approx [0.1,0.2]$, where the system exhibits bistable behavior (shaded region in Fig.~\ref{rosssp} (b)). Figure \ref{rosssp} (c) demonstrates the temporal bistable phenomena for $\epsilon = 0.15$, where oscillatory state and OD state may coexist. However, with enhancement of coupling parameter $\epsilon > 0.2$ (approx.), the system settles down into OD state. This feature is depicted through Fig.~\ref{rosssp} (d) for $\epsilon = 0.3$.

\begin{figure}[!t]
	\centerline{\includegraphics[width=0.3\textwidth]{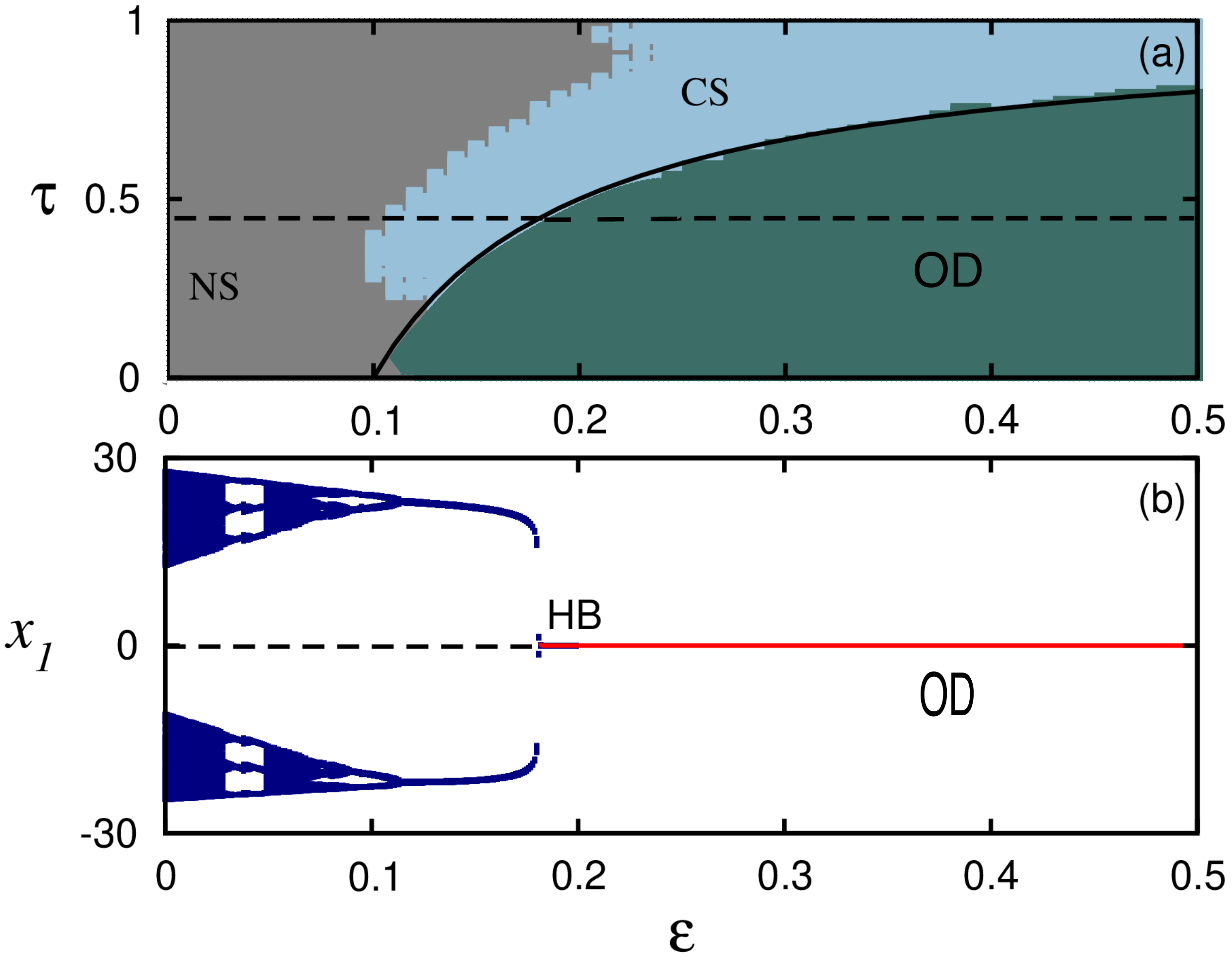}}
	\caption{  (a) {\bf Full $\epsilon$-$\tau$ phase-diagram for system \eqref{rosseq}}: Three different dynamical domains of $N=100$ coupled R\"ossler oscillators  are distinguished in the parameter plane $(\tau-\epsilon)$. The notation NS, CS and OD are same as given in Fig.\ \ref{rosssp}. (b) {\bf Bifurcation diagram against the coupling strength with fixed $\tau=0.45$}: The horizontal dashed line in subfigure (a) corresponds to $\tau=0.45$, for which this bifurcation diagram is explored. The other parameters are same as chosen in Fig.\ \ref{rosssp}. For the smaller values of coupling strength $\epsilon$, coupled system only exhibits the oscillatory behavior. For comparatively higher values of $\epsilon > 0.2$ (approx.), the coupled system is stabilized at coupling dependent stable steady state.}
	\label{rosstime}
\end{figure}
%

\par Similarly, we plot the phase diagram in the parameter plane $\tau-\epsilon$ for R\"ossler oscillator in Fig.~\ref{rosstime} (a) for the temporal interaction defined in Eq.\ (\ref{temporaleq3}). Proceeding to the way as proposed in Sec.\ \ref{TDI},  a time-dependent linear stability analysis is carried out. 
Linear stability analysis at the stationary point provides nontrivial characteristic equations,
\begin{eqnarray}
	\lambda =  \Big(\dfrac{a}{2}\Big) \tau  + (1-\tau) \Bigg(\frac{a-\epsilon}{2}\Bigg).
\label{rosscond}
\end{eqnarray}

Setting $\lambda = \alpha + i \gamma$, where $\alpha$ and $\gamma$ are real and imaginary parts of the eigenvalues, Eq.~(\ref{rosscond}) leads to 
\begin{eqnarray}
	\alpha = \tau  \epsilon + a -\epsilon.
\label{condross}
\end{eqnarray}

The solid black line in Fig.~(\ref{rosstime}) (a), corresponds to the locus $(\alpha = 0$ in  Eq.~(\ref{condross}))
\begin{eqnarray}
	\tau = 1-\frac{a}{\epsilon}, \hspace{0.2cm} \epsilon \ne 0.
\label{condtempros}
\end{eqnarray}
Analytical condition (solid black line) of Eq.\ (\ref{condtempros}) matches well with the numerically calculated OD region in Fig.~\ref{rosstime} (a). Here the dynamics changes from periodic attractor to a stationary point via Hopf bifurcation as a real part of the eigenvalue $\alpha$ becomes negative. Figure \ref{rosstime} (a) reveals a transition from incoherent (desynchronized) to coherent behavior happens depending on the value of $\tau$ and $\epsilon$. For comparatively lower values of $\tau$, the same transition occurs, but the system settles down from desynchronized oscillatory state to OD state. Our detailed eigenvalue analysis is also supported by bifurcation analyses in Fig.~\ref{rosstime} (b) for $\tau=0.45$, where it is shown that the dynamic coupling can induce a transition between periodic attractor and OD even in identical R\"ossler oscillators. 

\begin{figure}[!t]
	\centerline{\includegraphics[width=0.3\textwidth]{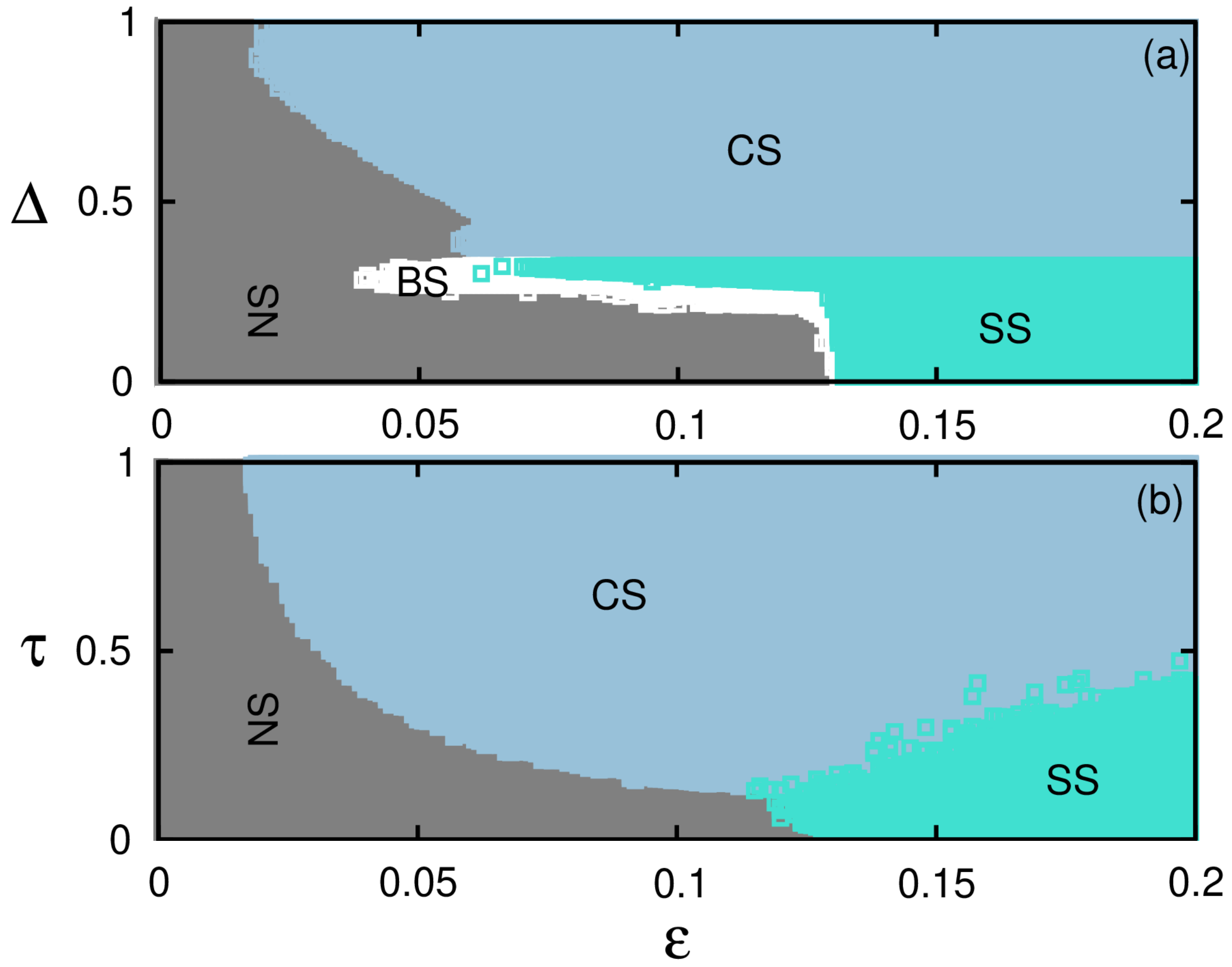}}
	\caption{ {\bf The phase diagram of Rosenzweig-MacArthur model in \textbf{(a)} ($\Delta, \epsilon$)  and \textbf{(b)} ($\tau, \epsilon$) at $T=34.6$}: Four different regimes are observed, including no synchronization (NS) (gray), complete synchronization (CS) (sky-blue), bi-stable state (BS) (white) and steady state (SS) (turquoise). The dynamics of coupled system changes from desynchronized state to the death state or the synchronized state depending on the control parameters $\Delta$ and $\tau$ as coupling strength $\epsilon$ is varied. Coexistence of two attractors (BS) are noticed over a narrow region of $\Delta-\epsilon$ in the subfigure (a).}
	\label{mcar}
\end{figure}

\subsection{Ecological System: Rosenzweig-MacArthur (RA) model}

\par In order to further exemplify the dynamic interaction, we consider the Rosenzweig-MacArthur (RA) model ~\cite{goldwyn2008can, rosenzweig1963graphical}, in which dynamics belonging to each of the patch of a meta-population is described by the following equations
{\begin{eqnarray}
	\hspace{2cm}	\dot{x_i} &=& rx_i\Bigg(1-\frac{x_i}{K}\Bigg) - \alpha_i \frac{x_i}{x_i+E}y_i + \epsilon (H\overline{x}-x_i),\nonumber \\
	\hspace{2cm}	\dot{y_i} &=& y_i\Bigg(\alpha_i \xi\frac{x_i}{x_i+E} - m\bigg),
	\end{eqnarray}
	where $x_i$ and $y_i$ are, respectively, vegetation and herbivore density, $r$ is intrinsic growth rate,
	$K$ is carrying capacity, $\alpha_i = \alpha$ is the maximum predation rate of the predator, $E$ is the half-saturation constant, $\xi$ represents predator efficiency, and $m$ is the mortality rate of the predator. The Rosenzweig-MacArthur model perhaps the simplest model that can actually be applied in real ecosystems. As a result, this model becomes a standard prey-predator model in consumer-resource dynamics in theoretical ecology. Here, we choose the parameter values as $r =0.5$, $K = 0.5$, $\alpha=1$, $E=0.16$, $\xi=0.5$, and $m=0.2$. Results for spatial and temporal dynamic interaction in parameter space $(\Delta, \epsilon)$ and $(\tau, \epsilon)$ are shown in Figs.\ \ref{mcar} (a) and \ref{mcar} (b), respectively. In Goldwyn and Hastings~\cite{goldwyn2011roles}, both the species can
	disperse between patches, but here for the sake of simplicity
	we consider dispersal of only one of the species (i.e.,
	vegetation) between patches. The numerical findings in Figs.~\ref{mcar} (a) and \ref{mcar} (b) attest the generic signature of our proposed dynamic interaction schemes. Just like the earlier numerical simulations, here we also able to portray the transition from desynchronized region (NS) to either synchronized regime (CS) or to death states (SS) depending on the interplay of $\epsilon$, $\Delta$ and $\tau$. Furthermore, a bistable region (BS) is found in Fig.~\ref{mcar} (a), where the oscillatory dynamics coexist with death states.

	\section{Conclusion and Summary}\label{conc}
	\par The overarching motivation of this work is to explore the effect of spatial as well as temporal dynamic interaction in coupled nonlinear oscillators. We have studied the transition to steady state in interacting nonlinear oscillators with effective dynamic mean-field interactions. Note that the focus is given on the interaction depending on the mean-field control parameter in the phase space instead of considering mobile agents configuration, where the nodes are moving in the phase space neglecting the oscillator's internal dynamics \cite{chowdhury2019extreme,9170822}. We have found that the spatial and temporal dynamic control parameters $\Delta$ and $\tau$ play a vital role in the transitions to various synchronization regions as well as in the suppression of oscillations. There is an optimal window in the parameter space of coupling strength and spatial dynamic control parameter, where the system undergoes to the steady state via first or second order transitions. Recently, the attention has been shifting away from continuous phase transition to first-order dynamical transition \cite{verma2019explosive,bi2014explosive,verma2017explosive,chen2013explosive,verma2019explosivePRE} due to its relevance in biological and chemical processes. However, our proposed dynamic interaction is a rare example of coupled system, which can offer fresh new insights due to its time-varying interaction strategy. And this important consideration may appeal at least few interested young researchers due to its applicability in various natural as well as social systems. We would like to explore the first-order dynamical transitions in detail under the limelight of phase transitions in near future. In temporal dynamic interaction, using an approximate linear stability analysis, we obtained the threshold values of the coupling strength for the transition to death state, and it is in agreement with the numerical results. The spatial interaction also brings discontinuous occasional temporal interaction, which is analyzed by calculating  $\tau_{avg}(\Delta)$ for a given $\Delta$. Analytic estimates are supplemented by numerics for several systems. Particularly, the threshold of coupling strength, which can facilitate death state of the whole system, is invariable with the change of network size. Spatial dynamic interaction in general display multiple asymptotic dynamical states in relation to different initial conditions due to the interplay of $\epsilon$ and $\Delta$ in a region of bistability. 
	The coexistence of such two phases (the oscillatory state and the steady state) at the same time is ubiquitous in many realistic scenarios including biological \cite{beuter2003nonlinear} and chemical systems \cite{aguda1990experimental}. Our account of presented results may yield new insights and foster the understanding of the temporal dynamics of coupled oscillators.
		\section*{Acknowledgment}
		MDS and AP acknowledges financial support (Grant No.\ EMR/2016/005561 and INT/RUS/RSF/P-18) from Department of Science and Technology (DST), Government of India, New Delhi. S.N.C. would  like  to  acknowledge  the  CSIR  (Project No. 09/093(0194)/2020-EMR-I) for financial assistance.	
	\section*{DATA AVAILABILITY}
	The data that support the findings of this study are available within the article.\\

\bibliographystyle{apsrev4-1}
\bibliography{biblo}

%


\end{document}